\documentstyle[epsfig]{mn}
\begin{document}
\LARGE
\normalsize

\title[GX 339-4]
{High resolution radio observations of the black hole candidate GX 339-4}

\author[R. P. Fender et al]
{R.~P.~Fender$^1$\thanks{E-mail: rpf@star.maps.susx.ac.uk}, 
R.~E.~Spencer$^2$, S.~J.~Newell$^2$ and A.~K.~Tzioumis$^3$\\
$^1$ Astronomy Centre, University of Sussex, Falmer, Brighton BN1 9QH, UK\\
$^2$ University of Manchester, Nuffield Radio Astronomy Laboratories,
Jodrell Bank, Macclesfield, Cheshire SK11 9DL, UK\\
$^3$ Australia Telescope National Facility, CSIRO, PO Box 76, Epping 2121,
NSW, Australia}

\maketitle

\begin{abstract}

We present radio observations of the black hole candidate X-ray binary
GX 339-4 with the Australia Telescope compact array. Mapping of the
highest resolution 3.5 cm data reveals a jet-like extension, which if
confirmed would be the first detection of a radio jet from a {\em persistent}
black-hole candidate system. No evidence is found for associated structures
such as bow shocks or jet lobes on larger scales. The spectral energy 
distribution from 22 - 3 cm is relatively flat, suggesting emission is
dominated by a compact absorbed core.

\end{abstract}


\section{Introduction}

Radio emission is an observational property of $\sim 15$\% of X-ray
binaries (see e.g. Hjellming \& Han 1995). It is anticorrelated with
the property of X-ray pulsations (Fender et al. 1997) and displays the
nonthermal spectra and high (T$_b > 10^8$ K) brightness temperatures
characteristic of synchrotron emission.

Radio jets from X-ray binaries were first discovered in the late 1970s
from SS 433 (Spencer 1979) in which the ejecta were subsequently
observed to have an ejection velocity of 0.26 c (Hjellming \& Johnston
1981).  Over the ensuing $\sim 15$ years radio-jets were discovered
from perhaps six further X-ray binary sources, in none of which
however accurate proper motions of individual ejecta, or plasmons,
could be measured (see e.g. Fender, Bell Burnell \& Waltman 1997 for a
review of observational properties). In the past three years however,
apparent superluminal plasmon velocities, corresponding to true
relativistic bulk motions at $\sim 0.9$c have been observed from two
transient X-ray binaries, GRS 1915+105 (Mirabel \& Rodriguez 1994) and
GRO J1655-40 (Tingay et al. 1995).

GX 339-4 is an unusual X-ray binary which exhibits distinct and varied
X-ray states, but is in all states persistently and clearly observable
with X-ray telescopes, unlike the transient radio-jet sources GRS
1915+105 and GRO J1655-40.  The characteristics of the X-ray emission,
notably a hard tail, rapid variability and a similarity to Cyg X-1,
have lead to classification of the source as a strong black hole
candidate (e.g. Makishima et al. 1986; Miyamoto et al. 1992; Miyamoto
et al. 1993; Nowak 1995).  Optical studies (Callanan et al. 1992)
reveal a 14.8 hr modulation, probably orbital in origin, in the light
from the 16-19th magnitude variable optical counterpart. Distance
estimates to the source range from 1.3 kpc (Predehl et al. 1991) to
$\geq 4$ kpc (Makishima et al. 1986).

The discovery of radio emission from the source was announced in Sood
\& Campbell-Wilson (1994), and sporadic monitoring has been ongoing
over the past $\sim 18$ months at 36 cm using the Molonglo Observatory
Synthesis Telescope (MOST).  The source was also detected during the
southern hemisphere radio survey of X-ray binaries conducted by
Spencer et al. (1997).  Durouchoux et al. (1997) and Sood et
al. (1997) discuss the radio behaviour of the source over the past
$\sim 18$ months in the context of X-ray monitoring and similarities
to Cyg X-1.

\section{Observations}

On 1996 July 11-14 we observed the source with the Australia Telescope
compact array (ATCA), in a high-resolution 6km
configuration. Observations were made simultaneously at 6.3 \& 3.5 cm
for the first three nights, and at 21.7 \& 12.7 cm on the fourth (in
order to obtain broad band spectral coverage). The source was clearly
detected on all nights and at all four wavelengths; the mean flux
densities at 21.7, 12.7, 6.3 \& 3.5 cm are listed in Table 1, and
plotted in Figure 1. 

GX 339-4 has a roughly flat radio spectrum, similar to that of Cyg X-3
in quiescence (e.g. Waltman et al. 1994), and indicative of a compact
absorbed core. A best fit single power-law (also indicated in Fig. 1)
has a spectral index ($\alpha = \Delta \log S / \Delta \log \nu$) of
$0.16 \pm 0.03$. The extension of this relatively flat spectrum to 36
cm was confirmed by MOST observations on 1996 July 13.5 (Hunstead,
private communication).  There were clear night-to-night variations in
the flux density and spectral index of emission at 6.3 \& 3.5 cm;
these will be discussed in the context of simultaneous
multi-instrument X-ray observations in a follow-up paper. There is no
evidence in the data for flux variations on timescales of an hour or
less of amplitude $>30$\%, unlike the rapid large-amplitude
quasi-periodic radio oscillations observed occasionally from GRS
1915+105 (Pooley 1996a, 1996b).  Judging by the flux monitoring
observations of Sood et al. (1997), the source was in a relatively
bright radio state at the time of these observations.

We have mapped the lowest-resolution 21.7 cm data (Figure 2) in order
to look for arcmin-scale structures such as the extended jets of
1E1740.7-2942 \& GRS 1758-258 (Mirabel 1994), the surrounding
synchrotron nebula of Cir X-1 (Stewart et al. 1991), or bowshocks such
as that reported by Hunstead, Wu \& Campbell-Wilson (1997) to be be
associated with the GRO J1655-40 jets.  No such structures are evident
in our map. Most sources detected at 21.7 and visible in Fig. 2 are
also in the 36 cm Molonglo maps of GX 339-4 in Durouchoux et
al. (1997) and Sood et al. (1997). The nearest reasonably strong
source to GX 339-4 lies 2 arcmin to the NE.

\begin{figure}
\centering
\leavevmode\epsfig{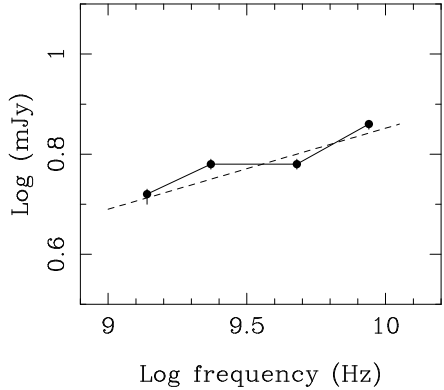}
\caption{22 -- 3 cm spectral energy distribution of GX 339-4. Also shown
is best-fit single power-law of spectral index $0.16 \pm 0.03$.}
\end{figure}

\begin{table}
\caption{Mean radio flux densities of GX 339-4 at 20, 13, 6 \& 3 cm
obtained with ATCA over 1996 July 11-14.}
\label{symbols}
\begin{tabular}{cc}
Wavelength (cm) & Flux density (mJy) \\
21.7 & $5.2 \pm 0.2$ \\
12.7 & $6.0 \pm 0.1$ \\
6.3  & $6.0 \pm 0.1$ \\ 
3.5  & $7.2 \pm 0.1$ \\
\end{tabular}
\end{table}

\begin{figure}
\leavevmode\epsfig{file=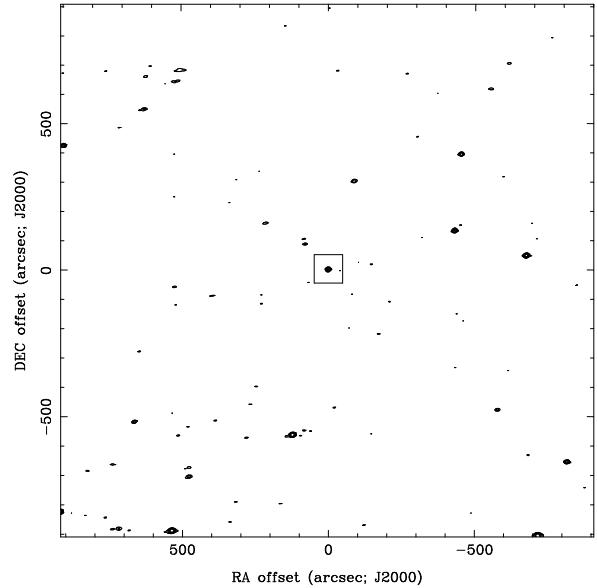, width=8cm, bbllx=24, bblly=150, 
bburx=571, bbury=682, clip}
\caption{30 $\times$ 30 arcmin 21.7 cm map of GX 339-4 region, convolved
with a 10 arcsec circular restoring beam. All sources more than 3$\sigma$
above the r.m.s. noise of $\sim 0.2$ mJy are shown. GX 339-4 is
indicated by the box; the vertices of the box are eight times longer than
the axes of Figure 2. There is no evidence of large-scale structure 
such as lobes, bow shocks or synchrotron nebulae
associated with GX 339-4. Most of the sources in the 36 cm Molonglo 
maps of Durouchoux et al. (1997) and Sood et al. (1997) are clearly
visible in this figure.}
\end{figure}

\begin{figure}
\leavevmode\epsfig{file=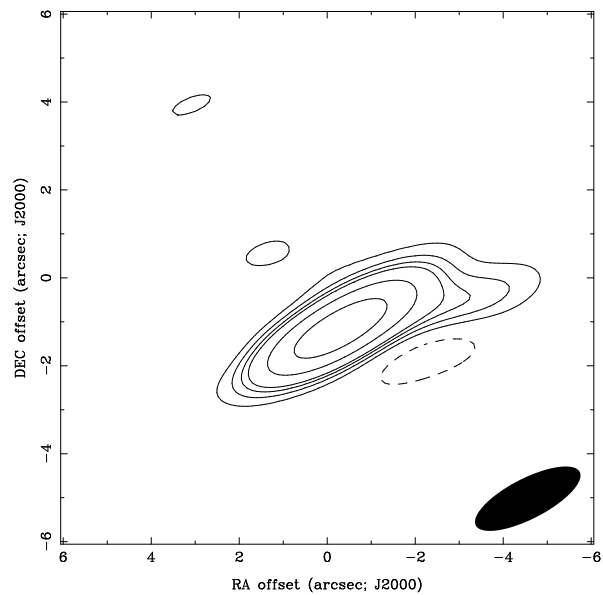, width=8cm, bbllx=24, bblly=150, 
bburx=571, bbury=682, clip}
\caption{3.5 cm super-uniform weighted 
map of GX 339-4 obtained with ATCA in high resolution 6km
configuration. Contours are at -5, 5, 10, 15, 20, 40 and 80 times
r.m.s. noise of 50 $\mu$Jy.  The jet-like feature extending to the
west of the source is detected at $\geq 15 \sigma$.  The solid ellipse
represents the synthesised beam.}
\end{figure}

However, the most intriguing aspect of the observations was the
detection of extended emission from the source in the mapping of the
highest-resolution 3.5 cm data (Fig 3).  This feature was first
apparent in preliminary mapping of the data at ATCA, and its existence
has subsequently been confirmed by re-mapping of the data both with
the MIRIAD software at Sussex University, and AIPS at Jodrell Bank.
Furthermore, the feature is present in maps deconvolved with both the
CLEAN algorithm and the Maximum Entropy procedure, and no similar
features appear in the images of any other objects mapped over the
period of our entire observing run from July 1 to July 14. However,
we cannot at this stage rule out an origin for the feature in 
atmospheric phase errors (though the phase solutions are good) as
the source is not strong enough for self-calibration.

The feature is significant at $\geq 15\sigma$ level, but there is no
evidence of a counterjet. Mapping of previous ATCA observations
obtained by Spencer et al. (1997), and Sood et al. (1997) reveals no
evidence for such a feature, implying a transient nature. The
structure cannot be resolved in our lower resolution longer wavelength
maps. Removal of a Gaussian fit to the core source reveals the plasmon
to lie $3 \pm 1$ arcsec to the west of GX 339-4, where uncertainties
in this process are reflected in the large error estimates.

\section{Discussion : a radio jet ?}

The lack of an observed counterjet places a lower limit on the ratio
of the brightnesses of the approaching and receding components of
$\geq 4$. Both the black-hole candidacy of GX 339-4 and the apparent
one-sidedness of the jet point to a relativistic ejection.  Under the
assumption of symmetric ejection of identical optically thin plasmons
at $\sim 0.9$c (as for GRS 1915+105 - Mirabel \& Rodriguez 1994), we
can place an upper limit on the angle to line of sight of the ejection
of $\sim 75^{\circ}$. Assuming the observed feature to be the
approaching plasmon, we can expect a lower limit on the observed
proper motion to be $\geq 0.2/d$(kpc) arcsec d$^{-1}$. However,
caution must be exercised in making these assumptions, in particular
as the spectral index of the ejecta is not clear, and ejections from
GRO J1655-40 can reveal an intrinsically asymmetric nature (Hjellming
\& Rupen 1995).  The lack of an observed jet-like structure to a
resolution of $\sim 1$ arcsec in observations made 313 days before
ours (data from Spencer et al. 1997), places a lower limit on the
velocity of the ejecta of $v \geq 6 \times 10^8 (d/kpc)$ cm
s$^{-1}$. At a distance of 1.3 kpc this corresponds to $\sim 0.03$c,
so even if not relativistic the outflow is clearly high velocity (but
at such a velocity it would need to be intrinsically one-sided as
relativistic beaming would not be important).

\begin{figure}
\leavevmode\epsfig{file=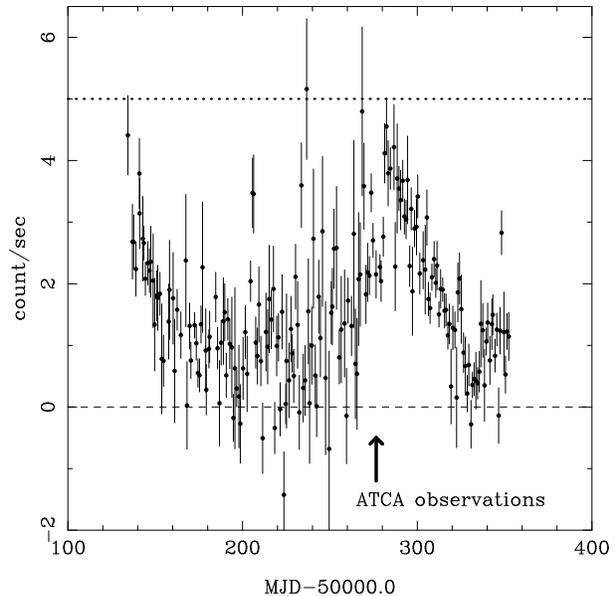, width=8cm}
\caption{XTE ASM soft X-ray monitoring of GX 339-4 
in the 2-10 keV band, illustrating an
outburst near the time of the ATCA detection of the radio jet. 
In this energy range, the Crab corresponds to about 75 count/sec.
All data points represent one-day averages.  The dotted line at
5 count/sec represents the approximate flux level below which the
source is considered to be `off', the lowest of its X-ray states
(Nowak 1995 and private communication).}
\end{figure}

Soft X-ray monitoring with the XTE ASM during the time of our
observations is shown in Fig 4. The source appears to be undergoing an
X-ray outburst at the time of the observations, although the overall
flux level corresponds to {\em off}, the lowest of the X-ray states
(Nowak 1995 and private communication).  A poorly-understood but clear
relation between X-ray and radio behaviour has already been
established for GRS 1915+105 and GRO J1655-40 (Foster et al. 1996;
Harmon et al. 1995).  Given the time required to achieve the observed
angular separation, if associated with the outburst the ejection would
have needed to occur around the beginning of the X-ray event, which
may be consistent with the behaviour of GRS 1915+105. Note that
Miyamoto \& Kitamoto (1991) have already predicted the existence of a
jet in GX 339-4, however this was in order to explain its X-ray
properties in the {\em very high} state.

\section{Conclusions}

GX 339-4 has become the latest black hole candidate X-ray binary to be
detected as a variable radio source. Durouchoux et al. (1997) and Sood
et al. (1997) have already discussed possible radio - X-ray
correlations in the source, and have compared it to Cygnus X-1, the
most famous persistent black hole candidate radio source. In four days
of radio observations with ATCA in high-resolution configuration we
have measured the spectral energy distribution of GX 339-4 from 22 - 3
cm and mapped the region of the source on scales from arcmin to
arcseconds. We find a relatively flat ($\alpha \sim +0.2$) radio
spectrum for the emission, indicative of a compact absorbed core.
Low-resolution mapping finds no evidence on arcmin-scales for lobes.
bow shocks or synchrotron nebulae such as those associated with other
radio-jet X-ray binaries.

Most intriguingly however we have presented evidence for the
existence of a jet-like structure in radio maps of GX 339-4. If
confirmed, this source becomes the first persistent black-hole
candidate source to reveal a radio jet.

The importance of the detection of a jet, which is in all likelihood
relativistic, from a non-transient black hole candidate X-ray
binary cannot be understated. Both GRS 1915+105 and GRS J1655-40
are transient sources, and while displaying quasi-recurrent 
behaviour in the 2-3 years since their discovery, the lack of
detection of these systems by previous X-ray satellites
(Lochner \& Whitlock 1992; Castro-Tirado 1994; 
Castro-Tirado, private communication) testifies
to prolonged states below detectable levels. GX 339-4 on the
other hand is a persistently visible source, both at X-rays
and (at least for past $\sim 18$ months) in the radio,
and may turn out to be one of our most valuable laboratories
in the study of radio jets.

\section*{Acknowledgements}

We are happy to acknowledge fruitful and stimulating discussions
with Ravi Sood, Philippe Durouchoux, Mike Nowak, Alberto 
Castro-Tirado, Bob Sault and Vince McIntyre. We
acknowledge quick-look results provided by the ASM/XTE team.

\end{document}